# Characteristics of solitary waves in a relativistic degenerate ion beam driven magneto plasma

*Manoj Kr. Deka[1#], Apul N. Dev[2], Amar P. Misra[3] and Nirab C. Adhikary[4]*

[1]Department of Applied Sciences, Institute of Science and Technology, Gauhati University, Guwahati- 781014, Assam, India

[2]Center for Applied Mathematics & computing, Siksha 'O' Anusandhan (Deemed to be University), Khandagiri, Bhubaneswar-751030, Odisha, India.

[3]Department of Mathematics, Siksha Bhavana, Visva-Bharati University, Santiniketan-731 235, India.

[4]Physical Sciences Division, Institute of Advanced Study in Science and Technology, Vigyan Path, Paschim Boragaon, Garchuk, Guwahati-781035, Assam, India

#Corresponding author E-mail:manojd143@gmail.com

Abstract:

The nonlinear propagation of small amplitude ion acoustic solitary wave in relativistic degenerate magneto plasma in presence of ion beam is investigated in detail. The nonlinear equations describing the evolution of solitary wave in presence of relativistic non-degenerate magnetized positive ions and ion beams including magnetized degenerate relativistic electrons are derived in terms of Zakharov-Kuznetsov (Z-K) equation for such plasma systems. The ion beams which are ubiquitous ingredient in such plasma systems, are found to have a decisive role in the propagation of solitary wave in such highly dense plasma system. The conditions of wave, propagating with typical solitonic characteristics are examined and discussed in detail under suitable conditions of different physical parameter. Both subsonic and supersonic wave can propagate in such plasmas bearing different characteristics under different physical situations. A detailed analysis of waves propagating in subsonic and/or supersonic regime is carried out. The ion beam concentrations, magnetic field as well as ion beam streaming velocity is found to play a momentous role on the control of the amplitude and width of small amplitude perturbation both in weakly(or non-relativistic) and relativistic plasmas.

**Introduction**

The investigations on the plasma wave spectra in degenerate as well as non-degenerate plasmas and its related features have been getting a boost during the last few years due to its prominent role in understanding the collective interactions in astrophysical systems affected by the relativistic degeneracy factor like white and brown dwarfs, magnetars, neutron stars as well as in laboratory experiments of intense laser-matter interaction[1-13]. On the other hand matter in extreme states has been well explored in laboratory where it is produced by intense ion beams and in fact these ion beams are regarded as the indispensable tools to reproduce plasma of the origin of neutron star, white dwarfs etc. in laboratory [14-22].It has been well established and reported in different literatures that these ion beams are also ubiquitous in space plasma environment submerged in magnetic field starting from plasma sheet boundary layers of the Earth's magnetosphere through magneto sphere of different planets [23] to supernova-driven plasma flows[24] along with in the environment of pulsars, blazars and quasi-perpendicular shock in the interstellar medium [25]where relativistic effects play an important role in the formation of forerunners of the ion-acoustic solitary waves. The importance of the ion beam-plasma interaction process in different plasma environment has been actively studied by the research fraternity throughout the world [26-30]. Recently few attempts have been made to study the salient features of relativistic degenerate plasma specially in the astrophysical plasma environment where such plasma conditions are prevalent. For example Esfandyari-Kalejahi et al studied the characteristics of solitary wave in degenerate electrons and positrons and found that solitary waves of both subsonic and supersonic range can propagate in ultra-relativistic degenerate plasmas whereas non-relativistic plasma supports only subsonic ion acoustic solitary wave. Moreover they also pointed out the possibility of propagation of Compressive solitary wave in non-relativistic and ultra-relativistic degenerate plasmas[31]. Masud and Mamun reported that the in degenerate plasma of astrophysical plasma environment, the constituent degenerate plasma ingredient specially the electron density is found to have a decisive in controlling the amplitude, width and speed of the solitary wave[32]. Behery et al discussed the stability and propagation of non-linear excitations for supersonic relativistic quantum plasmas by obtaining Zakharov-Kuznetzov-type equation and highlighted the possible application regarding from the view point of laboratory and space plasma[33]. El-Shamy et al studied the non-linear solitary wave propagation in dense degenerate magneto plasma and they concluded that ion cyclotron frequency, direction cosine of wave vector has a great impact on the propagation of solitary wave in such plasma[34]. Hossen et al studied the characteristics of ion acoustic shock waves in presence of heavy ions in degenerate plasmas in nonplanar geometry and found that the basic features such as speed, amplitude, width of the shock waves gets modified due to the degenerate pressure and number density of electron[35]. Rahman et al investigated the properties of solitary waves in relativistic degenerate dense plasma and concluded that compressive solitary wave propagate in different relativistic regime and the amplitude and width of the ion acoustic solitary pulses increases due to enhancement of relativistic factor by the increased number density of the plasma species[36].

Wang et al discussed the features of ion acoustic solitary waves in quantum plasmas with two polarity ions and relativistic electron beams and found that the presence of relativistic electron beam changes the dispersion relation in such plasma[37]. Mikaberidze and Berezhiani obtained stable solution of electromagnetic soliton for relativistic and nonrelativistic degenerate plasmas by applying relativistic hydrodynamics and Maxwell equations[38]. Rahman et al. studied, the dynamics of low frequency nonlinear ion wave in a weakly dissipative solitons in dense relativistic-degenerate plasmas[39]. Hossen and Mamun studied the propagation characteristics of electroacoustic waves in fully relativistic astrophysical degenerate plasmas and concluded that the relativistic degeneracy played a significant role in modifying the basic characteristics of solitary waves in such plasmas[40]. Shukla et al in their report discussed the nonlinear ion modes in relativistic degenerate plasmas in presence strong coupling of ions and pointed out the possibility of critical dependence of coupling and degeneracy on the basic features of the solitary wave[41]. Apart from solitary waves, the shock waves[42] as well as instability [43]due to amplitude modulation is also carried out in such degenerate plasmas. On the other hand, Irfan et al studied the nonlinear Magneto-acoustic solitary and shock waves with relativistic degenerate electrons and found that for a certain density of the relativistic electrons the wave frequency reaches the lowest possible value and then increases. Moreover a spatial reduction thereby strengthening the wave amplitude was also reported by them[44]. Mc Kerr et al investigated the conditions for occurrence as well as structural features of large amplitude nonlinear structures considering relativistically degenerate electrons and relativistically non-degenerate ions in astrophysical plasma environment and found that the amplitude of solitary pulses increases both with increased Mach number and also by moving the system towards more relativistic by increasing the density[45]. Here in this brief report, we have investigated the features of solitary waves in presence of magnetized relativistic ion and ion beams as well as relativistically degenerate electrons by deriving the dynamical evolution equation in terms of Zakharov-Kuznetsov (Z-K) equation which to the best of our knowledge and belief has never been addressed before. We feel this will open up a new exciting area of study where this type of plasma environment is of particular importance either from the view point of laboratory and space astrophysical plasmas.

## II. Theoretical Formulation

We consider the nonlinear propagation of small amplitude ion acoustic solitary wave in relativistic degenerate magneto plasma. The basic equations which include the equations for relativistic positive ions and ion beams

$$\partial_t n_{i,b} + \nabla \cdot (n_{i,b} \vec{v}_{i,b}) = 0 \qquad \text{---------- (1)}$$

$$(\partial_t + \vec{v}_{i,b} \cdot \nabla)(\gamma_{i,b} \vec{v}_{i,b}) = (e/m_{i,b})\{\vec{E} + (\vec{v}_{i,b}/c) \times \vec{B}\} \qquad \text{---------- (2)}$$

$$eE + (\vec{v}_e/c)B + (\gamma_e^2/\vec{n}_e)\{\vec{\nabla} p + (1/c^2)\partial_t P_e\} = 0 \qquad \text{---------- (3)}$$

$$\nabla \cdot \vec{E} = 4\pi e(n_e - n_i - n_b) \qquad \text{---------- (4)}$$

Where $P_e = \pi m_e^4 c^5 / 3h^3 \left[ s(2s^2 - 3)(s^2 + 1)^{1/2} + 3\sinh^{-1} s \right]$, $s = pF_e/m_e c$, is a dimensionless parameter and $pF_e = (3h^3 n_e/8\pi)^{1/3}$ is the momentum of the electrons on the Fermi surface where *h* is the Planck's constant. Assume that $n_e = n_{e0} + n_{e1}$ where $n_{e0}$ is the unperturbed number density for electrons, $n_{e1}$ is the perturbation of the electron number density $(n_{e1} \ll n_{e0})$. Using Taylor expansion technique [46] in Eq. (3),

$$P_e \simeq P_{e0} + (\partial_{n_e} P_e)_{n=n_{e0}} n_{e1} + (1/2)(\partial_{n_e^2}^2 P_e)_{n=n_{e0}} n_{e1}^2 \quad \text{or} \quad P_e \simeq P_{e0} + (2\varepsilon_{Fe} n_{e1}/3\gamma_0) + (s_0^2 + 2/9\gamma_0^3 n_{e0}) \varepsilon_{Fe} n_{e1}^2$$

With $\gamma_0 = \sqrt{1 + s_0^2}$, $s_0 = pF_e/m_e c$, $\varepsilon_{Fe} = (3\pi^2 n_e)^{2/3} \hbar^2 / 2m_e$, $\hbar = h/2\pi$, Where the relativistic factor is given by,

$\gamma_j = (1 - v_j^2/c^2)^{-\frac{1}{2}}$, $v_j^2 = v_{xj}^2 + v_{yj}^2 + v_{zj}^2$; $\gamma_j = 1 + v_j^2/2c^2$, $j(=e,i,b)$ and $E = -\nabla \varphi$. Equations (1) - (4) are normalized using the following normalized parameter...

$\phi = (\varepsilon_{Fe}/e)\varphi$, $t = \tau \omega_j^{-1}$, $x = X\lambda_{Fe}$, $N_j = n_j/n_{j0}$, $V_j = v_j/C_s$, Thus $\lambda_{Fe}\omega_j = C_s$, $\omega_j = (4\pi n_{j0} e^2/m_j)^{1/2}$,

$\lambda_{Fe} = (\varepsilon_{Fe}/4\pi n_{i0} e^2)^{1/2}$, $C_s = \sqrt{\varepsilon_{Fe}/m_i}$.

Thus the normalized set of equation becomes

$$\partial_T N_j + \nabla \cdot (N_j \times V_j) = 0 \quad \text{-----------(5)}$$

$$(\partial_T + V_j \times \nabla)\{\gamma_j(V_j)\} = -\nabla\varphi - V_j \times \Omega_{Bj} \quad \text{------------(6)}$$

$$\nabla\varphi = -V_e H + (\gamma_e^2/N_e)(\partial_X P_e + \alpha \partial_T P_e) \quad \text{------------(7)}$$

$$\nabla^2\varphi = (\mu_e N_e - N_b \mu_b - N_i), \quad \mu_e = n_{e0}/n_{i0}, \quad \mu_b = n_{b0}/n_{i0} \quad \text{------------(8)}$$

Where $H = B_0 \big/ c(4\pi n_{i0} e^2 m_i)^{1/2}$, $\Omega_{Bj} = eB_0/\omega_j m_j c$, j=(i-ion,b-beam), $\alpha = m_e(\gamma_0 - 1)/m_i$, $\gamma_0 = \sqrt{1+s_0^2}$

### III. Derivation of Z-K equation

In order to investigate the propagation of ion acoustic solitary wave, we use the standard reductive perturbation technique to derive the Z-K equation for the present plasma model under consideration. We use the following stretched Co-ordinate $\xi = \varepsilon^{1/2}(X - \lambda\tau)$, $\eta = \varepsilon^{1/2}Y$, $\zeta = \varepsilon^{1/2}Z$, $\tau = \varepsilon^{3/2}T$, Where $\lambda$ is the Phase velocity. The plasma parameters can be expanded as, $N_\alpha = 1 + \sum_{j=1}^{\infty}\varepsilon^j N_\alpha^{(j)}$, $\varphi = \sum_{j=1}^{\infty}\varepsilon^j \varphi^{(j)}$, $V_{bx} = V_{b0} + \sum_{j=1}^{\infty}\varepsilon^j V_b^{(j)}$, $V_{ex,ix} = \sum_{j=1}^{\infty}\varepsilon^j V_{ex,ix}^{(j)}$, $V_{by,z} = \sum_{j=3/2,k=1}^{\infty}\varepsilon^j V_{by,z}^{(k)}$, $V_{iy,z} = \sum_{j=3/2,k=1}^{\infty}\varepsilon^j V_{iy,z}^{(k)}$, $V_{ey,z} = \sum_{j=3/2,k=1}^{\infty}\varepsilon^j V_{ey,z}^{(k)}$. Substituting the respective expansions through equations (5) to (8) and equating the lowest order of $\varepsilon$, we get $V_{ix}^{(1)} = \varphi^{(1)}/\lambda$, $N_i^{(1)} = \varphi^{(1)}/\lambda^2$, $V_{bx}^{(1)} = \varphi^{(1)}/(\lambda - V_{b0})\gamma_{b1}$, $N_b^{(1)} = \varphi^{(1)}/(\lambda - V_{b0})^2 \gamma_{b1}$, $N_e^{(1)} = \varphi^{(1)}/(\alpha_1 + 2\beta_1)$ and the dispersion relation for the nonlinear wave speed given by $\lambda^2 = \{(\alpha_1 + 2\beta_1)/\mu_e\}\{(\mu_b \mu_m/\rho^2 \gamma_{b1}) + 1\}$ with $\rho = (1-\delta)$, $\delta = V_{b0}/\lambda$. Together with $\partial_\eta \varphi^{(1)} + \partial_\zeta \varphi^{(1)} = -\Omega_{Bb}(V_{bz}^{(1)} + V_{by}^{(1)})$, $\partial_\eta \varphi^{(1)} + \partial_\zeta \varphi^{(1)} = -\Omega_{Bi}(V_{iz}^{(1)} + V_{iy}^{(1)})$ and

$\mu_e N_e^{(1)} = \mu_b N_b^{(1)} + N_i^{(1)}$. Now equating the next higher order of $\varepsilon$ we get

$$\partial_\tau N_{ix}^{(1)} - \lambda\partial_\xi N_{ix}^{(2)} + \partial_\xi V_{ix}^{(2)} + \partial_\xi(N_{ix}^{(1)}V_{ix}^{(1)}) + \partial_\eta V_{iy}^{(2)} + \partial_\zeta V_{iz}^{(2)} = 0 \quad \text{-------------(9)}$$

$$\partial_\tau N_{bx}^{(1)} - (\lambda - V_{b0})\partial_\xi N_{bx}^{(2)} + \partial_\xi V_{bx}^{(2)} + \partial_\xi(N_{bx}^{(1)}V_{bx}^{(1)}) + \partial_\eta(V_{by}^{(2)}) + \partial_\zeta V_{bz}^{(2)} = 0 \quad \text{------------(10)}$$

$$\partial_\tau V_{ix}^{(1)} - \lambda\partial_\xi V_{ix}^{(2)} + V_{ix}^{(1)}\partial_\xi V_{ix}^{(1)} + \partial_\xi \varphi^{(2)} = 0, \quad \text{------------(11)}$$

$$\gamma_{b1}\partial_\tau V_{bx}^{(1)} - \{(\lambda - V_{b0})/V_{b0}\}\gamma_{b2}\partial_\xi V_{bx}^{(1)^2} - (\lambda - V_{b0})\gamma_{b1}\partial_\xi V_{bx}^{(2)} + \gamma_{b1}V_{bx}^{(1)}\partial_\xi V_{bx}^{(1)} + \partial_\xi \varphi^{(2)} = 0 \quad \text{-------------(12)}$$

Where $\gamma_{b1} = (1 + 1.5\gamma^2)$, $\gamma_{b2} = 1.5\gamma^2$, $\gamma = V_{b0}/c$

$$(\partial_\xi \varphi^{(2)} + N_{ex}^{(1)}\partial_\xi \varphi^{(1)}) = (\alpha_1 - \alpha_2\lambda)\partial_\xi N_{ex}^{(2)} + 2(\beta_1 - \beta_2\lambda)N_{ex}^{(1)}\partial_\xi N_{ex}^{(1)} + (\alpha_2 + \beta_2)\partial_\tau N_{ex}^{(1)} \quad \text{------------(13)}$$

$$\partial_\eta V_{iy}^{(2)} + \partial_\zeta V_{iz}^{(2)} = (\lambda/\Omega_{Bi}^2)\partial_\xi(\partial_{\eta^2}^2 \varphi^{(1)} + \partial_{\zeta^2}^2 \varphi^{(1)}) \text{ and } \partial_\eta V_{by}^{(2)} + \partial_\zeta V_{bz}^{(2)} = ((\lambda - V_{b0})/\Omega_{Bb}^2)\partial_\xi(\partial_{\eta^2}^2 \varphi^{(1)} + \partial_{\zeta^2}^2 \varphi^{(1)})$$

Along with $\partial^2_{\xi^2}\varphi^{(1)} + \partial^2_{\eta^2}\varphi^{(1)} + \partial^2_{\zeta^2}\varphi^{(1)} = \mu_e N_e^{(2)} - \mu_b N_b^{(2)} - N_i^{(2)}$ ----------(14)

Eliminating second order term from Eqs. (9) - (14) and with the help of first order equations, we obtain the final form Z-K equation as follows

$$\partial_\tau \varphi^{(1)} + A\varphi^{(1)}\partial_\xi \phi^{(1)} + B\partial^3_{\xi^3}\varphi^{(1)} + C\partial_\xi \left(\partial^2_{\xi^2}\varphi^{(1)} + \partial^2_{\zeta^2}\varphi^{(1)}\right) = 0;$$
$$A = q/p;\ B = 1/p;\ C = r/p$$
------------(15)

Where

$$p = \left[\{\mu_e(\alpha_2 + \beta_2)/(\alpha_1 - \alpha_2\lambda)(\alpha_1 + 2\beta_1)\} + \{2\mu_b\gamma_{b1}/(\lambda - V_{b0})^3 \gamma_{b1}^2\} + (2/\lambda^3)\right]$$

$$q = \begin{bmatrix} \{2\mu_e(\beta_1 - \beta_2\lambda)/(\alpha_1 - \alpha_2\lambda)(\alpha_1 + 2\beta_1)^2\} - \{\mu_e/\{(\alpha_1 + 2\beta_1)(\alpha_1 - \alpha_2\lambda)\}\} - \\ \{\mu_b\{2\gamma_{b2}(\lambda - V_{b0}) - 3\gamma_{b1}\}/(\lambda - V_{b0})^4 \gamma_{b1}^3\} + 3/\lambda^4 \end{bmatrix}$$

$$r = 1 + \mu_b/\Omega_{Bb}^2 + 1/\Omega_{Bi}^2$$

Where $\alpha_1 = 2/3\gamma_0$, $\beta_1 = (s_0^2 + 2)/9\gamma_0^3 n_{e0}$, $\alpha_2 = 2\alpha/3\gamma_0$, $\beta_2 = ((s_0^2 + 2)\alpha)/9\gamma_0^3 n_{e0}$

Now to solve the Z-K equation we used the transformation $\chi = \gamma(l\xi + m\eta + n\zeta - U\tau)$ and considering $\varphi^{(1)}(\xi,\eta,\zeta,\tau) = \psi(\chi)$ which gives, $-U\psi + (Al/2)\psi^2 + \gamma^2 l\left(Bl^2 + C(m^2 + n^2)\right)d^2_{\chi^2}\psi = 0$. To derive the required solution of Z-K Eq. (15) we used the well-known *tanh*-method and for that the transformation $z = \tanh(\chi)$ and $\psi(\chi) = w(z)$ is introduced and then the equation above becomes

$$-Uw + (Al/2)w^2 + \gamma^2 l\left(Bl^2 + C(m^2 + n^2)\right)\left((1-z^2)^2 d^2_{z^2}w - 2z(1-z^2)d_z w\right) = 0$$ ----------(16)

For finding the series solution of Eq. (16) substituting $w(z) = \sum_{r=0}^{\infty} a_r z^{k+r}$ and for leading order analysis of finite terms gives $r = 2$ and $k = 0$ and then the $w(z)$ becomes $w(z) = a_0 + a_1 z + a_2 z^2$. Now substituting the value of $w(z)$ in Eq. (17) $w(z) = a_0 + a_1 z + a_2 z^2 = a_0(1 - z^2)$, where $a_0 = -a_2$ and $a_1 = 0$ we get the stationary solution of Z-K equation (15) as

$$\varphi = \varphi_m \sec h^2\{\chi/W\}$$ ---------- (17)

where $\varphi_m = 3U/Al$ and $W = \frac{1}{2}\left[U/l\left(Bl^2 + C(1-l^2)\right)\right]^{1/2}$ are the amplitude and width of the solitary wave solution respectively.

**IV. Results and Discussion.**

The effects of various plasma parameters which is of the interest of the present investigation on the propagation characteristics of the solitary wave as well as the basic features of the solitary waves in different regime of operation of plasma are examined analytically on the basis of the stationary solutions of the evolution equation given by (17). Here all the background plasma parameters are adopted as described by Zhenni et. al. and Misra and Shukla[46,47]. Here since we have used the pressure equation in full form, so we have the freedom of choosing any density corresponding to $S_0<1$ or $S_0>1$ for the case of weakly (or non-relativistic) and relativistic case respectively. Magnetic fields are in the range of $10^9$ G ~ $10^{10}$ G, Positive ions are basically assumed to be Helium which is most abundant in such plasma environment. Initial streaming velocity of beam ions are so chosen that $\gamma$ is always less than unity.

Fig 1 describes the variation of phase velocity of the plasma system with the variation of relativistic streaming ion beam velocity for degenerate plasma within the above mentioned plasma parameter both in the case of weakly (or non-relativistic) and relativistic case for which $S_0<1$ and $S_0>1$ respectively. Here, as seen in Fig1(a), the variation of normalized phase velocity $\lambda$ with respect to relativistic beam velocity for three different concentration of ion beams in weakly or non-relativistic plasma is presented. Here it is clear from the figure that the phase velocity of the plasma system shows an enhancement with increasing ion beam concentration. This is because as we know that the presence of ion beam always makes a system more energetic which in turn moves the system with greater energy and hence with an increased phase velocity. On the other hand as evident from Fig.1(b), with increasing ion beam concentration in relativistic plasma, the phase velocity of the plasma system increases with an increase in the ion beam concentration which can be understood in the same context of the explanation provided for Fig1(a). However as evident from both the figures, the phase velocity shows a reduction in its value in the relativistic case than its value in the weakly or non-relativistic case. Basically here we have sketched the variation of phase velocity for two different regime of plasma operation namely weakly or non-relativistic and relativistic characterized by the dimensionless parameter $S_0<1$ and $S_0>1$ respectively. These two parameters are basically the function of plasma density. Here we have found $S_0<1$ for plasma density in the range of $10^{27}$ cm$^{-3}$ or less than that and $S_0>1$ for plasma density in the range of $10^{31}$ cm$^{-3}$ or greater than that. Clearly for the case of $S_0>1$, the plasma density is relatively high where with increasing ion beam concentration, the system can undergo more as well as frequent collision with energetic ion beams and thereby show a possible reduction in the phase velocity of the wave. On the other hand, it is evident from both the figures that with the present plasma parameters being considered for investigation, we can have both subsonic and supersonic regime wave propagation.

The variations of Non-linear co-efficient $A$ with ion beam concentration in the supersonic wave propagation regime for both the cases of both weakly relativistic and relativistic plasma is shown in Figure 2. Here as shown in both the figures the non-linear co-efficient decreases with increasing ion beam concentration in either of the case. Physically this can be attributed from the fact that ion beams which are an additional ingredient of the plasma system, adds to the non-linearity of the plasma system and as and when its concentration increases, which means if its concentration becomes similar to that like other ingredient (electron, ion etc.), will help the system to behave in a less non-linear way or otherwise the plasma system becomes somewhat uniform type in terms of three ingredients and thus the non-linearity of the plasma decreases or otherwise the plasma system will be somewhat like linear with the three ingredients . On the other hand, in the inset of Figure 2, the variation of the non-linearity of the plasma system with the relativistic wave propagation factor, which is also a function ion beam streaming velocity, is portrayed. And as seen from the figure, the non-linearity of the system gets enhanced with increasing relativistic factor or the ion beam streaming velocity. The prime reason behind this phenomenon is that with increasing initial streaming of ion beams(or otherwise relativistic factor) which is an additional parameter apart from the active plasma ingredients and whose prime role is associated with increasing the velocity of the plasma species, will always make the system to behave with an increased non-linearity. The variation of the non-linearity with initial ion beam streaming is carried out for weakly or non-relativistic case, however for the relativistic case also, same variation is noticed(not presented here).Here the other plasma parameters are same as discussed in case of Figure 1.

Figure 3 represents the simultaneous variation of non-linearity and dispersion in case of weakly or non-relativistic plasma ($S_0<1$) in the subsonic regime and it is clearly seen that there is almost a nice balance between the increased(decreased) non-linearity(dispersion).This balance is also seen in the supersonic wave propagation in case of non-relativistic and relativistic propagation. However interestingly this balance between non-linearity and dispersion is seen in the subsonic regime of wave propagation in case of weakly (or non-relativistic) plasma only and it is not observed in the subsonic regime of relativistic plasma(will be shown later on) though the variation of non-linearity and dispersion with ion beam concentration is more or less of similar nature. The reason behind this phenomenon may be understood from the fact that in the subsonic regime in weakly(or non-relativistic) plasma as the wave speed is smaller, the interaction or collision of the plasma wave with the ion beams are supposed to be prolonged which forces the system to behave in a rather non-linear way. On the other hand in case of relativistic plasma which is characterized by a relatively higher concentration of the plasma species for the present plasma system, the non-linearity of the plasma system increases considerably due to the same reason as discussed above. However the balance of nonlinearity and dispersion as observed in case of weakly or non-relativistic plasma(as described in Figure 3) is not noticed in case of relativistic plasma (not presented here). We have calculated the percentage of increase(decrease) of non-linearity(dispersion) in both non-relativistic and relativistic  increase case and found that with increasing ion beam concentration, the non-linearity of the plasma system increases by approximately by 36% whereas the dispersion is decreased by approximately 34% in subsonic wave propagation in the non-relativistic case, while in the relativistic case, the non-linearity of the plasma system increases by approximately by 50% whereas the dispersion is decreased by approximately 30% .The consequences of this peculiar observation is clearly observed while studying the variation of solitary structures which are to be discussed in the later part of the manuscript.

Figure 4 (a) and (b) describes the variation of the potential of the solitary wave with increased ion beam concentration both in the case of $S_0<1$ and $S_0>1$ respectively. As seen from the figure, in both the cases the solitary wave propagate with a nice balance of non-linearity and dispersion bearing the typical soliton characteristics. It is also clear from the figure that as the ion beam concentration increases, the balance of non-linearity and dispersion drives the wave with increasing amplitude which resembles to the typical variation of wave phase velocity as portrayed in Fig.1(a) and (b) as we can conclude that with increasing the phase velocity, the amplitude of the soliton will also increase thereby its width will decrease. This can also be understood from the typical variation of non-linearity of the plasma system with increasing ion beam concentration described in Fig2. Also for the present plasma model, the maximum amplitude as seen from the stationary solution of the evolution equation, is controlled by basically the variation of non-linear co-efficient. So, as the non-linearity increases , the amplitude decreases and vice-versa. At the same time the wave dispersion increases or decreases accordingly and thus the soliton characteristics of the plasma system remain intact. This is what is seen in Fig. 4(a) and (b) respectively.

On the other hand, Figure 5(a),and (b), represents the variation of soliton potential in weakly(non-relativistic) and relativistic plasma with ion beam concentration in the subsonic regime of wave propagation . As seen from the figure, the wave potential decreases with increasing ion beam concentration in both the cases, however the typical soliton characteristics is typically followed in case of subsonic regime of non-relativistic case whereas in the relativistic case the amplitude and width of the solitary wave decreases simultaneously. This can be understood from the typical variation of non-linearity and dispersion in such plasma as discussed in detail in case of Figure 3. Also as described in case of Figure 3, with prolonged collision , the wave phase velocity gets reduced due to the damping caused by the collision. Thus as ion beam concentration increases, the possibility of collision and thereby the damping of the phase velocity may occur which is the phenomenon behind the reduction of wave amplitude with increasing ion beam concentration. Also as for the present plasma system, the variation in soliton amplitude is controlled by the variation of the nature of non-linear co-efficient, so as the non-linearity increases (decreases),the soliton amplitude decreases(increases) which is clear from Figure 3.

Figure 6 represents the variation of solitary wave structure for two different value of the magnetic field in the subsonic regime of wave propagation for weakly relativistic case whereas the same is represented in the inset of the figure for the relativistic case. In both the cases it is clearly observed that the magnetic field has no control on the amplitude of the solitary wave whereas the width of the solitary wave gets enhanced with an increase in the strength of the magnetic field, i.e. the wave is found to be less dispersive with increasing the strength of the magnetic field. This may be because with such high range of magnetic field, due to greater confinement of the plasma species, the wave motion seems to be somewhat channelized in the direction of the magnetic field due to alignment of the plasma species with the direction of the magnetic field which forces the wave to be less dispersive as amplitude remains unaffected. The same is seen in case of Figure 7 also where the variation of the solitary wave in the supersonic range of the non-relativistic plasma is sketched and in the inset, the variation of the solitary wave is plotted with magnetic field in the relativistic plasma. Interestingly, the wave is the least dispersive for higher magnetic field in the supersonic regime of the relativistic plasma. This may be due to the fact that as in the supersonic regime, the wave phase velocity is larger compared to the subsonic regime, and the faster plasma species has a greater chance of being magnetized due to the synchronization of the cyclotron orbits of the plasma species with the magnetic field lines and also the relatively high plasma density allows a better confinement and hence it forces the system to be least dispersive.

Figure 8 represents the variation of solitary wave potential with increasing the relativistic streaming factor or relativistic streaming ion beam velocity. Here initial streaming velocity is so chosen that $\gamma$ is always less than unity. Here as seen from the figure, the amplitude of the solitary wave decreases with increasing the ion beam streaming velocity and at the same time its width also decreases. If we look at the detail inside of the underlying physics behind this phenomenon, we see that the streaming beam velocity which makes the system more and more non-linear, will attribute to the reduction of the amplitude of the solitary wave which can be attributed from the discussion of the nature of variation of non-linear co-efficient with equilibrium ion streaming energy sketched in the inset of Figure 2 which in turn controls the magnitude of the potential of the solitary wave. We can also conclude the nature of variation of the plasma wave phenomenon for the present plasma environment under discussion, with ion beam streaming velocity in linear and non-linear regime of wave propagation is different whereas it is more or less identical with ion beam concentration and relativistic streaming factor. Here basically we have discussed the nature of propagation of solitary wave in presence of relativistic ion beam streaming velocity in a degenerate relativistic plasma under the best possible and physically admissible range of which we believe will open up an exciting area of research in the ion beam plasma interaction in such highly dense, degenerate relativistic plasma with more detailed and better treatment of the physical parameter.

**V. Conclusion:**

Here in this brief report we have tried to depict salient features of relativistic degenerate plasmas in presence relativistic non-degenerate ion and ion beams in different regime of operation namely weakly or non-relativistic and relativistic. Both subsonic and supersonic waves are found to be supported by such type of plasma environment. The wave phase velocity with ion beam concentration reduces as we move from non-relativistic to relativistic plasma regime. Interestingly, in the non-linear analysis, the soliton amplitude variation with ion beam concentration closely resembles to the typical wave phase velocity variation in the linear regime. The variation of phase velocity both in subsonic and supersonic range is found to be pertinent with ion beam concentration. Typical soliton characteristics of the plasma waves are observed for different physical situation throughout the study except in the subsonic regime of relativistic plasma. The dominant effect of magnetic field in controlling the plasma dispersion in either subsonic or supersonic wave propagation in both non-relativistic and relativistic plasma is observed .Here we have tried to describe basic features of solitary wave propagation in presence of the above mentioned plasma ingredients analytically considering different possible physical situation. However the authors feel that a detailed numerical analysis of the present plasma system would certainly reveal many interesting features of solitary wave propagation in the above mentioned plasma systems though this is beyond the scope of the manuscript.

**Acknowledgement:**

One of the author, Manoj Kumar Deka would like to thank Department of Science and Technology, Govt of India, for providing financial assistance to carry out the research under a sanction by Science and Engineering Research Board via the Grant No. YSS/2015/001896.

**References.**

[1] N. C. Adhikary, M. K. Deka, and H. Bailung, Phys. Plasmas 16, 063701 (2009).
[2] A. N. Dev, J. Sarma , M. K. Deka and N. C. Adhikary, Plasma Sci. and Tech. **17**, 268 (2015) .
[3] A. N. Dev, J. Sarma  and M. K. Deka, Can. J. Phys. **93**, 1030 (2015)
[4] M. K. Deka, Braz. J. Phys.  **46**, 672 (2015)
[5] A. Mushtaq and S.V. Vladimirov,Eur. Phys. J. D **64**, 419 (2011)
[6] M. I.Shaukat, Eur. Phys. J. Plus **132,** 210(2017)
[7] A. Atteya, E. E. Behery and W. F. El-Taibany, Eur. Phys. J. Plus **132,**109(2017)
[8] N.L. Tsintsadze, L.N. Tsintsadze, A. Hussain and G. Murtaza, Eur. Phys. J. D **64**, 447 (2011)
[9] M. Mahdavi and F. Khodadadi Azadboni, Eur. Phys. J. D **68** 260, (2014)
[10] W. Masood, B. Eliasson,and P. K. Shukla, Physical Review E **81**, 066401 (2010)
[11] A.A. Mamun, P.K. Shukla Physics Letters A **374,** 4238 (2010).
[12] A.P. Misra, B. Sahu, Physica A **421,** 269 (2015).
[13] A. A. Mamun and P. K. Shukla,EPL, **94**, 65002 (2011).
[14] P.K. Shukla, and B. Eliasson, Rev. Mod. Phys. **83,** 885 (2011).
[15] R. Redmer, and G. Röpke, Contrib. Plasma Phys. **50**, 970 (2010).
[16] V. E. Fortov, D. H. H. Hoffmann, B Yu Sharkov, Physics – Uspekhi **51**,109 (2008).
[17] D.H.H. Hoffmann, A. Blazevic, P. NI, O. Rosmez, M. Roth, N.A. Tahir, A. Tauschwitz, S. Udrea, D. Varentsov, K.Weyrich,and Y. Maron, Laser and Particle Beams, **23**, 47(2005).
[18] A. Rahman  , S. A. Khan  and A. Qamar, Plasma Science and Technology, **17**,1000(2015).
[19] M. T. Gabdullin, S. K. Kodanova, T. S. Ramazanov, M. K. Issanova, T.N. Ismagambetova,  NUKLEONIKA ,**61**,125(2016).
[20] V.E. Fortov, *Extreme States of Matter on Earth and in the Cosmos,* (Springer-Verlag, Berlin Heidelberg, 2011)
[21] B.Y.Sharkov. D. H.H. Hoffmann. A. A. Golubev. Y. Zhao, Matter and Radiation at Extremes,**1**,28(2016).
[22] V.E.Fortov *Extreme States of Matter High Energy Density Physics*,(Springer Series in Materials Science,2016).
[23] K.Sauer and E. Dubinin Phys. Scr.,**T107**,167(2004)
[24] M. E. Dieckmann, A. Meli, P. K. Shukla, L. O. C. Drury and A. Mastichiadis, Plasma Phys. Control. Fusion **50,** 065020 (2008).
[25] M. E. Dieckmann, B. Eliasson, and P. K. Shukla, Physical Review E **70**, 036401 (2004).
[26] D.H.H. Hoffmann, N.A. Tahir, S. Udrea, O. Rosmej, C.V. Meister1, D, Varentsov,M. Roth, G. Schaumann, A. Frank, A. Blazevic, J. Ling, A. Hug, J. Menzel,Th. Hessling, K. Harres,M.G¨unther, S. El-Moussati1, D. Schumacher, and M. Imran, Contrib. Plasma Phys. **50**,7(2010)
[27] M. K. Deka, N. C. Adhikary, A. P. Misra, H. Bailung, and Y. Nakamura, Phys. Plasmas 19, 103704 (2012).
[28] C.-R. Choi, C.-M. Ryu, K.-C. Rha, K.-W. Min, and D.-Y. Lee,Physics of Plasmas 19, 032105 (2012).
[29] P. Chatterjee and R. Roychoudhury,Physics of Plasmas 2, 1352 (1995).
[30] K. K. Mondal, S. N .Paul and A. Roychowdhury, IEEE Trans. Plasma Sci. 26, 987 ( 1998)
[31] Esfandyari-Kalejahi, M. Akbari-Moghanjoughi, and E. Saberian, Plasma and Fusion Research **5**, 045 (2011).
[32] M. M. Masud and  A. A. Mamun, PRAMANA Indian journal of Physics **81**, 169 (2013).
[33] E. E. Behery,  F. Haas, and I. Kourakis, Physical Review E **93**, 023206 (2016).
[34] E. F. El-Shamy ,R. C. Al-Chouikh, A. El-Depsy,and N. S. Al-Wadie, Physics of Plasmas **23**, 122122 (2016).
[35] M.R. Hossen, L. Nahar, S. Sultana, A.A. Mamun, High Energy Density Physics **13,**13 (2014).
[36] A. Rahman, W. Masood, B. Eliasson and A. Qamar, Physics Plasmas **20,** 092304 (2013).
[37] Y. Wang, Y. Dong and B. Eliasson, Physics Letters A **377**,2604 (2013).
[38] G. Mikaberidze, V.I. Berezhiani, Physics Letters A **379**, 2730 (2015).
[39] A. Rahman, A. Mushtaq, S. Ali,  and A. Qamar, Commun. Theor. Phys. **59,** 479(2013).
[40] M. A. Hossen, and A. A. Mamun , IEEE Trans. Plasma Sci. 44, 643 (2016)
[41] P. K. Shukla  A. A. Mamun D. A. Mendis, Physical Review E **84**, 026405 (2011).
[42] H. R. Pakzad Canadian Journal of Physics, **89** 961( 2011).
[43] A. S. Bains, A. P. Misra, N. S. Saini,and T. S. Gill, Physics of Plasmas **17**, 012103 (2010).
[44] M. Irfan, S. Ali and Arshad M. Mirza, J. Plasma Phys. **82**, 905820106 (2016),
[45] M. M. Kerr, F. Haas and I. Kourakis, Physical Review E **90**, 033112 (2014).
[46] Z. Zhenni , W. Zhengwei , L. Chunhua and Y. Weihong ,Plasma Science and Technology **16**, 995(2014).
[47] A. P. Misra, and P. K. Shukla, Physical Review E **85**, 026409 (2012).

**Figures Captions:**

Fig1: (a) Variation of phase velocity with $\gamma$ in the case of $S_0 <1$ (b) Variation of phase velocity with $\gamma$ for the case of $S_0 >1$

Fig2. Variation of Non-linear Co-efficient $A$ with $\mu_b$ for the case $S_0 <1$ and $S_0 >1$ in the supersonic regime. Inset: Variation of Non-linear Co-efficient $A$ with $\gamma$ for the case $S_0 <1$.

Fig3. Variation of Non-linear Co-efficient $A$ and dispersive co-efficient $B$ with $\mu_b$ for the case $S_0 <1$ in the subsonic regime.

Fig4: (a) Variation of $\varphi$ of soliton with $\mu_b$ in weakly(or non-relativistic case for which $S_0<1$) in the supersonic regime (b) Variation of $\varphi$ with $\mu_b$ in the relativistic case for which $S_0<1$ in the supersonic regime.

Fig5: (a) Variation of $\varphi$ of soliton with $\mu_b$ in weakly(or non-relativistic case for which $S_0<1$) in the subsonic regime (b) Variation of $\varphi$ with $\mu_b$ in the relativistic case for which $S_0<1$ in the subsonic regime.

Fig6: (a) Variation of $\varphi$ with magnetic field in weakly or non- relativistic case in the subsonic regime: Inset variation of $\varphi$ with magnetic field in relativistic case in the subsonic regime.

Fig7: (a) Variation of $\varphi$ with magnetic field in weakly or non- relativistic case in the supersonic regime (b) Inset variation of $\varphi$ with magnetic field in relativistic case in the supersonic regime.

Fig8: Variation of $\varphi$ with $\gamma$ in weakly or non- relativistic case.

**Figures**

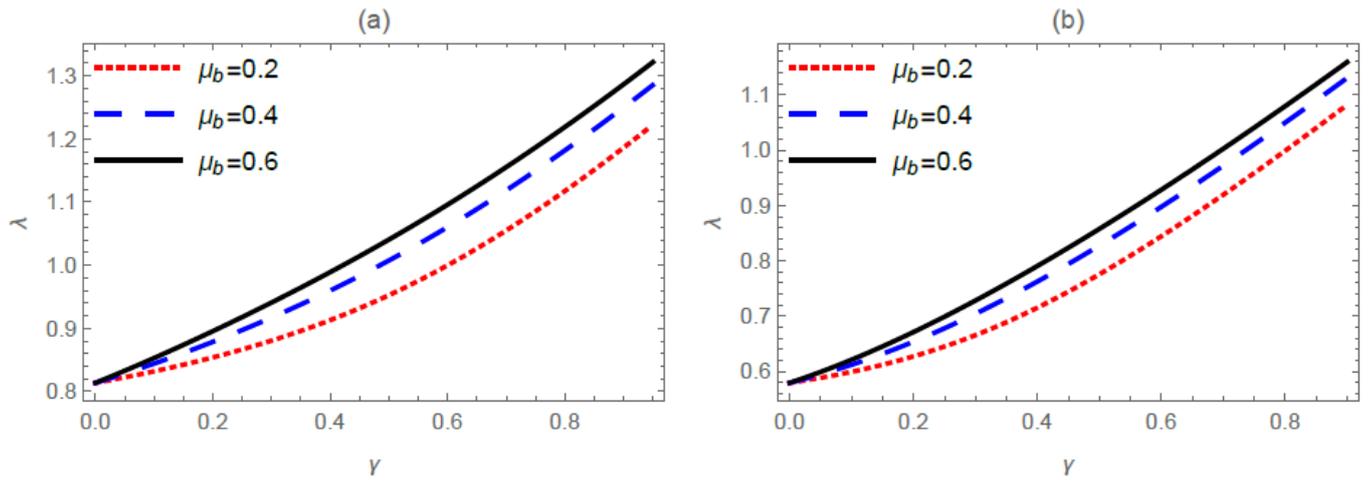

Fig1: (a) Variation of phase velocity with $\gamma$ in the case of $S_0 <1$ (b) Variation of phase velocity with $\gamma$ for the case of $S_0 >1$

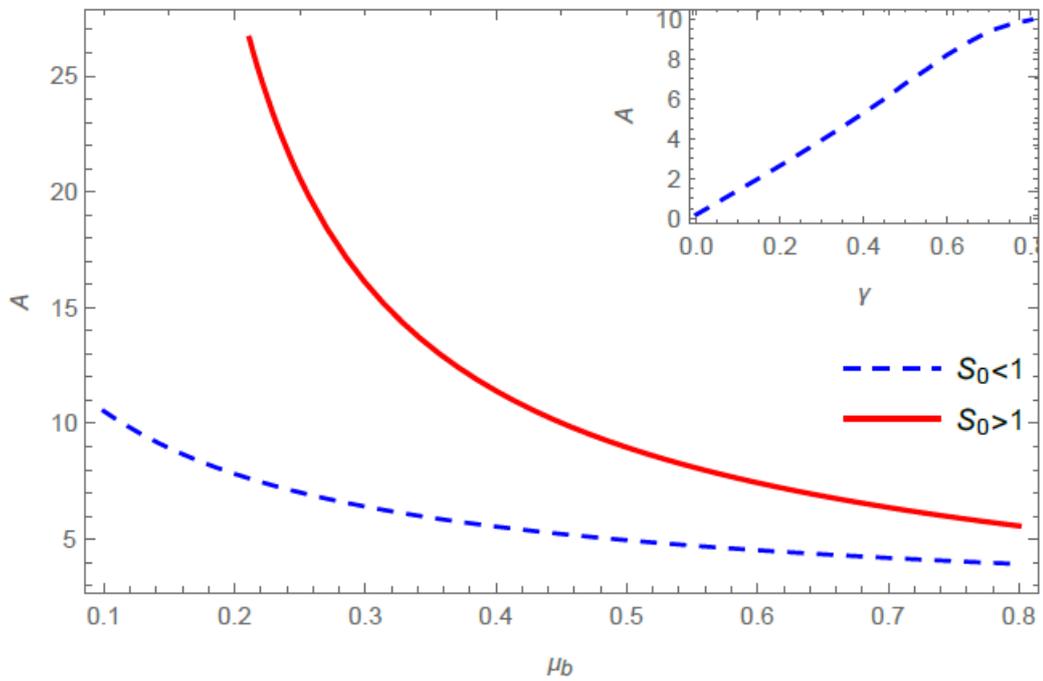

Fig2. Variation of Non-linear Co-efficient $A$ with $\mu_b$ for the case $S_0 <1$ and $S_0 >1$ in the supersonic regime. Inset: Variation of Non-linear Co-efficient $A$ with $\gamma$ for the case $S_0 <1$.

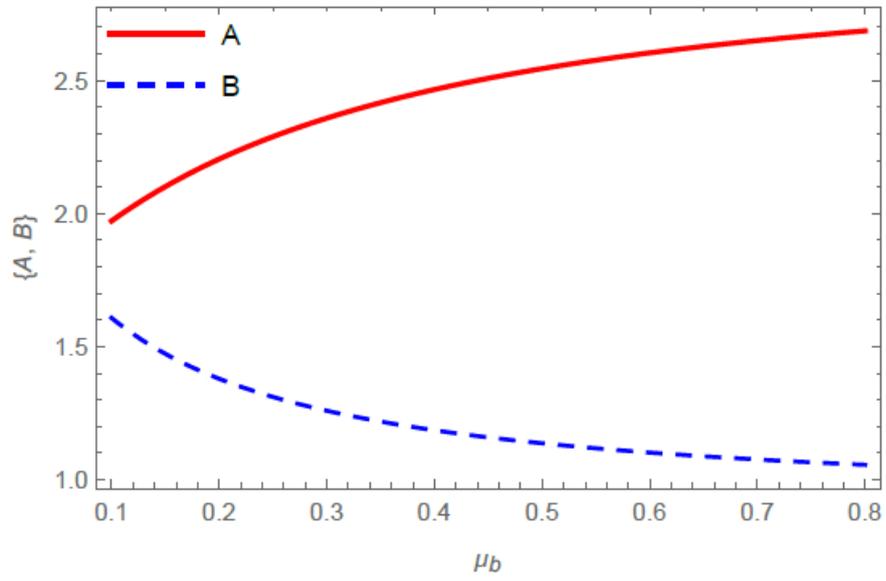

Fig3. Variation of Non-linear Co-efficient *A* and dispersive co-efficient *B* with $\mu_b$ for the case $S_0 <1$ in the subsonic regime.

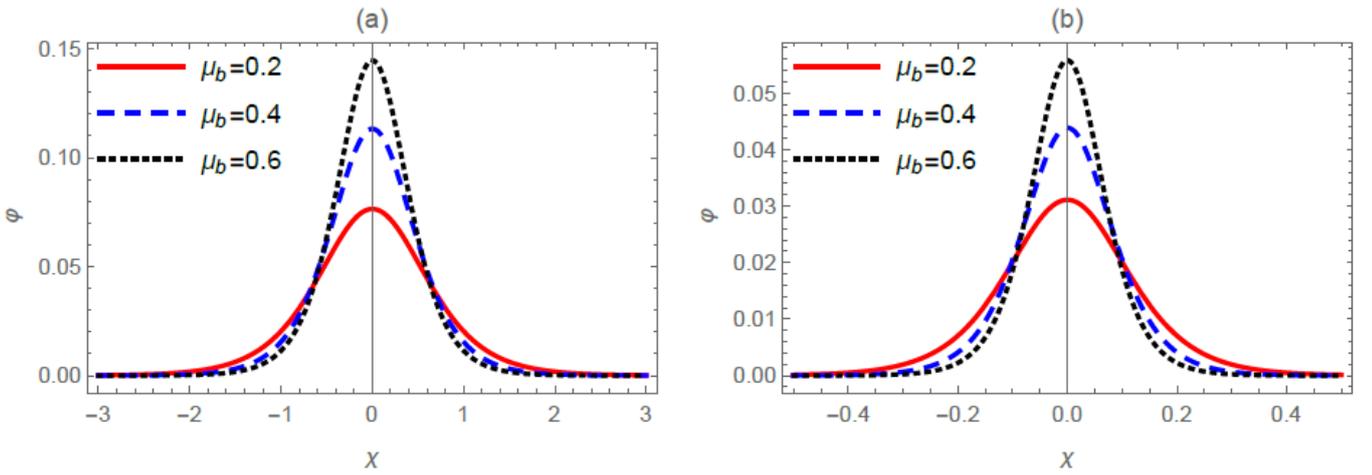

Fig4: (a) Variation of $\varphi$ of soliton with $\mu_b$ in weakly(or non-relativistic case for which $S_0<1$) in the supersonic regime (b) Variation of $\varphi$ with $\mu_b$ in the relativistic case for which $S_0<1$ in the supersonic regime.

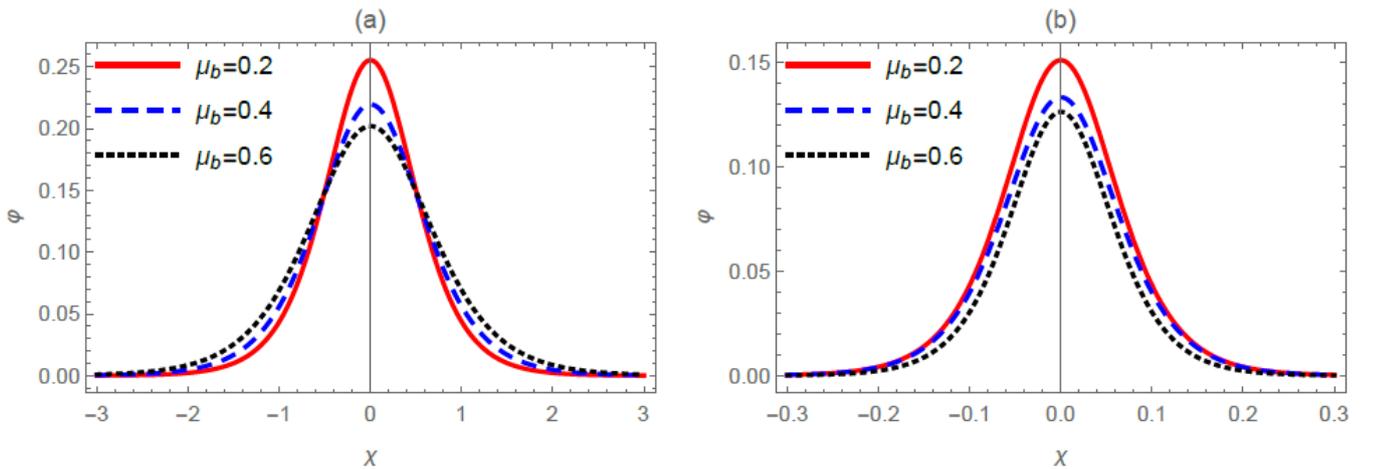

Fig5: (a) Variation of $\varphi$ of soliton with $\mu_b$ in weakly(or non-relativistic case for which $S_0<1$) in the subsonic regime (b) Variation of $\varphi$ with $\mu_b$ in the relativistic case for which $S_0<1$ in the subsonic regime.

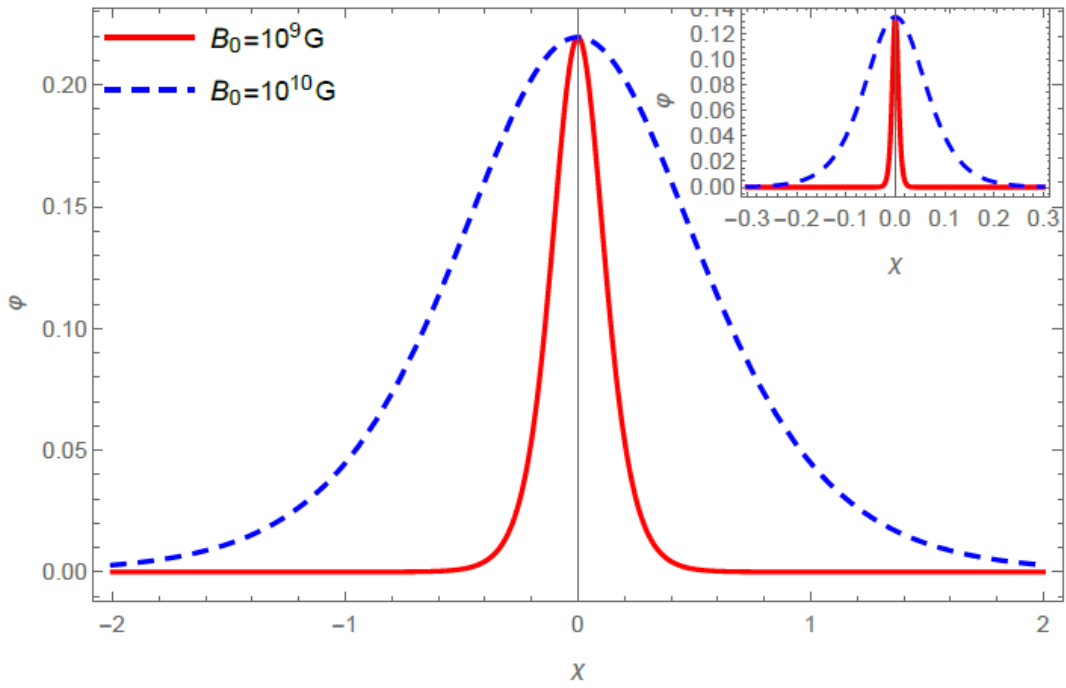

Fig6: Variation of $\varphi$ with magnetic field in weakly or non- relativistic case in the subsonic regime: Inset variation of $\varphi$ with magnetic field in relativistic case in the subsonic regime.

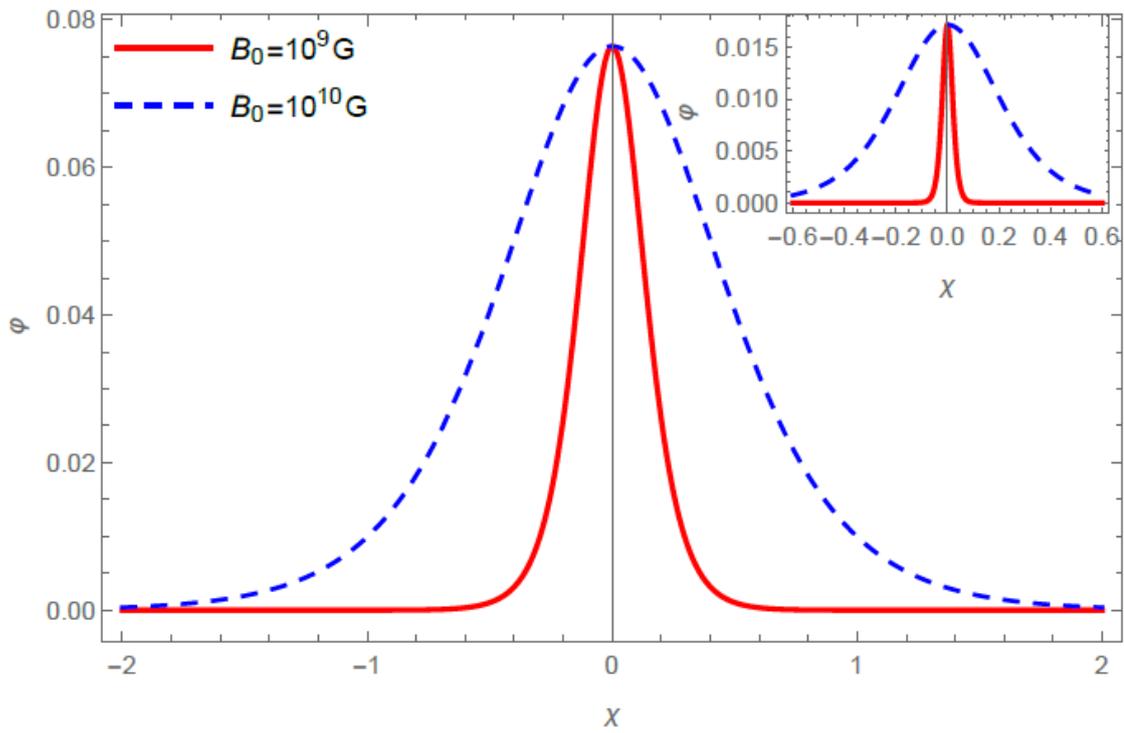

Fig7: Variation of $\varphi$ with magnetic field in weakly or non- relativistic case in the supersonic regime, Inset variation of $\varphi$ with magnetic field in relativistic case in the supersonic regime.

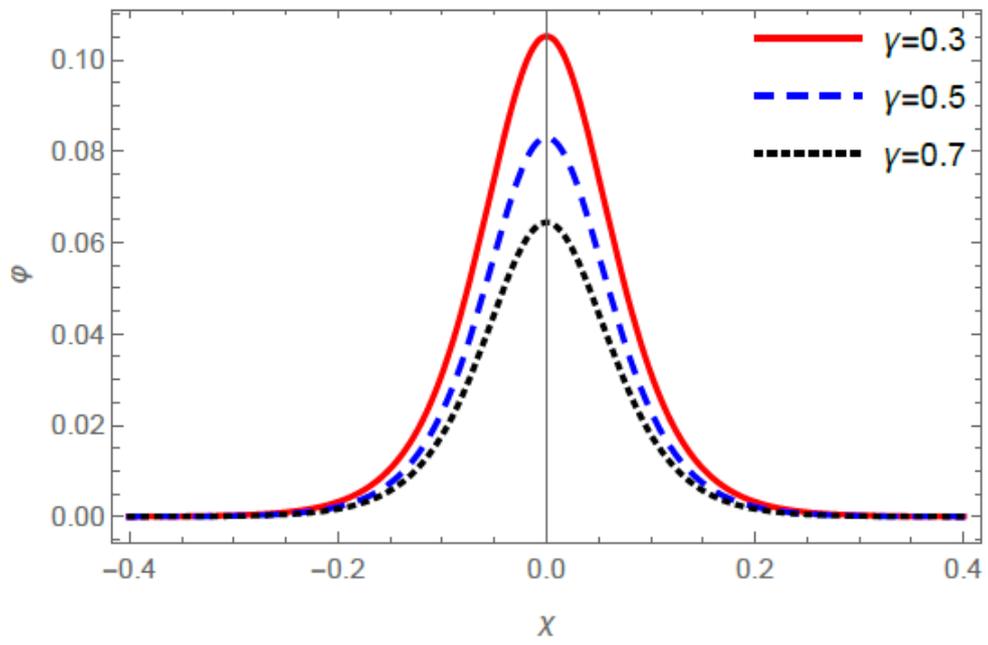

Fig8: Variation of φ with γ in weakly or non- relativistic case.